\documentclass[aps, english,superscriptaddress]{revtex4-2}

\usepackage[a4paper, margin=2cm]{geometry}
\usepackage[skip=5pt plus1pt, indent=10pt]{parskip}
\usepackage{mathptmx}

\usepackage[T1]{fontenc}
\usepackage[utf8]{inputenc}
\usepackage{booktabs}
\usepackage{threeparttable}

\usepackage[T1]{fontenc}
\newcommand{\model}[1]{\textsc{#1}}

\usepackage[normalem]{ulem}

\usepackage{graphicx}

\usepackage{hyperref}
\hypersetup{
  colorlinks,
  citecolor=blue,
  linkcolor=blue,
  urlcolor=blue,
}
\usepackage{siunitx}
\DeclareSIUnit\bar{bar}
\DeclareSIUnit\angstrom{\text {Å}}
\DeclareSIUnit\bohr{\text {b}}
\DeclareSIUnit\calorie{cal}

\usepackage{textgreek}
\usepackage{multirow}
\usepackage[version=3]{mhchem}

\usepackage{xcolor}

\begin{document}

\title{Overcoming sampling limitations using machine-learned interatomic potentials: the case of water-in-salt electrolytes}

\author{Luca Brugnoli}
\affiliation{Sorbonne Université, Muséum National d’Histoire Naturelle, UMR CNRS 7590, Institut de Minéralogie, de Physique des Matériaux et de Cosmochimie, IMPMC, F-75005 Paris, France}

\author{Mathieu Salanne}
\affiliation{Sorbonne Universit\'e, CNRS, Physicochimie des \'Electrolytes et Nanosyst\`emes Interfaciaux, F-75005 Paris, France}
\affiliation{R\'eseau sur le Stockage Electrochimique de l'Energie (RS2E), FR CNRS 3459, 80039 Amiens Cedex, France}
\affiliation{CNRS@CREATE, Singapore 138601, Singapore}

\author{A. Marco Saitta}
\affiliation{Laboratoire de Physique de L’\'{E}cole Normale Sup\'{e}rieure de Paris, CNRS, ENS \& Universit\'{e} PSL, Sorbonne Universit\'{e}, Universit\'{e} de Paris, 75005 Paris, France}

\author{Alessandra Serva}
\email{alessandra.serva@sorbonne-universite.fr}
\affiliation{Sorbonne Universit\'e, CNRS, Physicochimie des \'Electrolytes et Nanosyst\`emes Interfaciaux, F-75005 Paris, France}
\affiliation{R\'eseau sur le Stockage Electrochimique de l'Energie (RS2E), FR CNRS 3459, 80039 Amiens Cedex, France}

\author{Arthur France-Lanord}
\email{arthur.france-lanord@cnrs.fr}
\affiliation{Sorbonne Université, Muséum National d’Histoire Naturelle, UMR CNRS 7590, Institut de Minéralogie, de Physique des Matériaux et de Cosmochimie, IMPMC, F-75005 Paris, France}

\begin{abstract}
Machine-learned interatomic potentials hold the promise to enable the modeling of highly concentrated liquids over meaningful timescales, far from reach for current \textit{ab initio} electronic structure methods. Here we evaluate the performances of various MACE potentials in modeling a $21 m$ water-in-salt electrolyte based on lithium bis(trifluoromethanesulfonyl)imide. We test out-of-the-box foundation models, as well as both fine tuning and from scratch training strategies. Our simulations demonstrate that surrogate models allow to overcome sampling limitations of \textit{ab initio} molecular dynamics, reaching an excellent agreement with experimental observables such as the structure factor. We also demonstrate the benefit of fine tuning a foundation model over training from scratch: in terms of data efficiency, but most importantly as a means to provide information regarding configurations hard to sample, such as short Li$^+$--Li$^+$ distances. Finally, we show that depending on the reference exchange-correlation functional, empirical dispersion correction schemes can be detrimental. All in all, our work shows that machine-learned interatomic potentials are a good fit for the modeling of highly concentrated electrolytes over long timescales. 
\end{abstract}

\maketitle

\section{Introduction}

Aqueous electrolytes offer the potential to reduce the environmental toll of lithium-ion batteries and to significantly increase their safety regarding fire hazards. Unfortunately, common aqueous electrolytes degrade upon the application of voltages larger than water's electrochemical stability window (1.23 V). Water-in-salt electrolytes~\cite{Suo2015WiSE} (WiSEs) demonstrated early on their ability in overcoming this limitation. These are concentrated solutions in which there is more salt than water both in terms of volume and mass. The first device relying on a WiSE, which demonstrated stability up to 3 V, made use of a $21 m$ aqueous solution of a hydrophilic salt, lithium bis(trifluoromethanesulfonyl)imide (LiTFSI). This mixture has become the canonical example allowing to extend the electrochemical stability window of aqueous lithium electrolytes. At such high concentration, this mixture significantly departs from the ideal, dilute solution: ion pairing and solvent depletion generate a heterogeneous liquid where water-rich and anion-rich regions have been identified from both atomistic simulations and scattering experiments.\cite{Borodin2017Nanoheterogeneity,Horwitz2021Nanostructure} Overall, the highly concentrated LiTFSI aqueous electrolyte has become a staple for understanding solvation, ion transport mechanisms, and intermediate-range ordering in WiSEs. In addition, these phenomena have been related to properties such as structure, diffusion, viscosity, and conductivity, using numerical simulations such as molecular dynamics (MD)~\cite{ZhangMaginn2021WiSE1,ZhangMaginn2021WiSE2}. 

However, a consensus has not been reached yet regarding either the Li$^+$ solvation partitioning between water and bis(trifluoromethanesulfonyl)imide (TFSI$^-$), or the length scale of structural heterogeneity. In particular, estimates of these quantities are not fully consistent across different simulation or characterization techniques~\cite{Borodin2017Nanoheterogeneity,Horwitz2021Nanostructure}. Recently developed phenomenological models, based on ion crowding, clustering, and gelation, both in the bulk~\cite{mceldrewTheoryIonAggregation2020,mceldrewIonClustersNetworks2021} or at an interface~\cite{McEldrew2018WiSEdoublelayer} have been successfully reproducing trends observed in experimental or simulation setups, such as changes in activity as a function of salt concentration. They therefore allow to rationalize such phenomena in terms of simple model ingredients. Phenomenological models rely however on parameters, whose values must be constrained using molecular simulations. 

Classical MD simulations have been largely employed to investigate LiTFSI WiSEs from a microscopic perspective, probing phenomena such as Li$^+$ coordination, contact and solvent-separated ion pairs, as well as phase separation at the nanoscale, which correlates with sluggish solvent dynamics~\cite{ZhangMaginn2021WiSE1,Borodin2017Nanoheterogeneity}. Ion transport and viscosity~\cite{ZhangMaginn2021WiSE2,Li2019transport} have also been investigated, as sufficiently long timescales, in the range of tens of nanoseconds, can be reached using classical empirical potentials. Although inexpensive to use, these interaction models have some downsides: first, their empirical nature does not warrant any transferability. In essence, some properties could be in good agreement with experimental findings, while others could show large disagreement~\cite{Jinnouchi2025MLFFElectrochemistry}. Second, these models rely on static topological elements (bonds, angles, dihedrals), which prevent the investigation of reactivity. This is a showstopper if one wishes for instance to investigate the degradation of WiSEs under a voltage bias or in alkaline conditions~\cite{France-Lanord2022chemical}. 

These limitations motivate benchmarks on reference data. \textit{Ab initio} MD (AIMD) simulations based on density functional theory (DFT) provide such a reference; with appropriately validated exchange-correlation functionals, these simulations yield meaningful results for the structural and dynamical properties of molecular liquids. They show excellent transferability and enable the modeling of reactivity: in that sense, they allow going beyond empirical potentials. For LiTFSI WiSE, AIMD has been used as a means to probe solvation patterns~\cite{Malaspina2023unraveling,Wrobel2024ab,Goloviznina2024Nanochains,Singh2025structure}, spectroscopic signatures~\cite{Wrobel2024ab}, nanoscopic structuration~\cite{Goloviznina2024Nanochains}, diffusivity~\cite{Singh2025structure}, and intermediate-range order through the evaluation of the structure factor~\cite{Singh2025structure}. Unfortunately, the unfavorable algorithmic complexity of quantum chemistry methods prevents the application of AIMD to large system sizes or to long timescales. This is particularly important for LiTFSI WiSE: large simulation boxes are needed to probe intermediate- and long-range order, and the high viscosity and slow ion exchange dynamics require the sampling of trajectories over extended timescales. These limitations lead to simulations performed on too small simulation boxes, for too short simulation times, or using parameters controlling simulation accuracy set to too loose values. As we will discuss later on, we argue that the short simulation time bottleneck can be appreciated when attempting to compare computed and experimental structure factors~\cite{Singh2025structure}: there is significant disagreement in features at low wavenumber due to reorganization processes too slow to probe. 

Dependence on the method -- either the electronic structure method itself, or in the narrower context of DFT, the choice of exchange-correlation functional -- is nontrivial. Generalized gradient approximation (GGA) functionals and meta-GGAs differ in their description of hydrogen bonding and ion pairing, and the treatment of dispersion can change equilibrium density and ionic association propensities~\cite{Furness2020r2SCAN}. In this context, \citet{Goloviznina2024Nanochains} reported polymer-like Li--Li nanochains with short Li--Li bonds in BLYP-D3 simulations of a 128-pair cell, emphasizing that rare motifs can appear and must be assessed against reference choices and finite-size effects. Whether such motifs persist under alternative functionals, whether they survive longer sampling, and how they impact structure factors and transport remain open questions. These uncertainties complicate the use of AIMD alone as a production tool for dynamics properties, but they make it a valuable anchor point for training and validating interatomic potentials -- if the reference model is documented. 

In this context, machine-learned interatomic potentials (MLIPs) aim to reproduce a chosen reference model at a fraction of its computational cost, enabling nanosecond-to-microsecond sampling while retaining a direct link to electronic-structure labels~\cite{Jinnouchi2025MLFFElectrochemistry}. Their reliability is bounded by data coverage: configurations that are rare in short AIMD, such as short distances of same-charge ions, or large density fluctuations, can dominate extrapolation behavior in long trajectories. Model expressivity also constitutes a bound on the accuracy MLIPs can reach on reference data. While short-range interactions are well captured using modern architectures~\cite{Jacobs2025practical}, long-range physics are only starting to be considered and incorporated into MLIPs~\cite{Anstine2023machine,Grasselli2026long}. Since finite-cutoff models do not automatically recover dispersion or electrostatic tails, which influence properties such as density, compressibility, and dipole correlation functions, correction strategies should be included and validated following the same lines as the MLIP itself. 

MLIPs have been used a few times to model WiSEs. Wang \textit{et al.}~\cite{wangSwitchingRedoxLevels2023} trained deep potentials~\cite{Zhang2018deep} on LiTFSI aqueous solutions (up to $25.6 m$), to investigate how concentration affects redox levels, and ultimately the electrochemical stability window. More recently, Okuno~\cite{Okuno2025structures} used moment tensor potentials~\cite{Shapeev2016moment} to model WiSEs with two different salts, including LiTFSI; the focus was on solvation structure and exchanges using simulations at fixed density, as well as diffusion. Cao \textit{et al.}~\cite{Cao2025resolving} used deep potentials to investigate the solvation and transport properties of an aqueous ZnCl$_2$ electrolyte with variable concentration, reaching up to 30m. Overall, many key questions remain: are MLIPs able to reproduce experimental properties such as density, diffusion coefficients, as well as the structure factor? Are results obtained from typically short AIMD simulations biased due to an insufficient sampling of this highly viscous system? What does that entail regarding generating an appropriate dataset for MLIP training?

Here, we are attempting to answer these questions by focusing on a particular class of MLIPs: MACE, an O(3)-equivariant message-passing architecture that incorporates high body-order correlations and shallow networks, which has been shown to reach low force errors~ and high transferability\cite{Batatia2022MACE}. Beyond single-system, application-focused training, the MACE architecture has been used a series of so-called foundation models~\cite{Batatia2025Foundation,MACEFoundationsGitHub}: pre-trained models that cover a broad range of elements and are released in named families trained on large DFT datasets at specified levels of theory. These models can be used directly as baselines, or adapted by fine-tuning to a target chemistry of interest and reference data -- potentially using a different electronic structure method. This allows us to investigate the dependence on small, correlated datasets extracted from AIMD, and to explore if relying on an initial foundation model, either fine-tuned or used out-of-the-box, leads to significantly different behaviors compared to models obtained by training from-scratch. This is particularly interesting for WiSEs, since validation is not necessarily directly accessible through AIMD, due to the high viscosity of the system. 

\begin{figure}[h!]
    \centering
    \includegraphics[width=\textwidth]{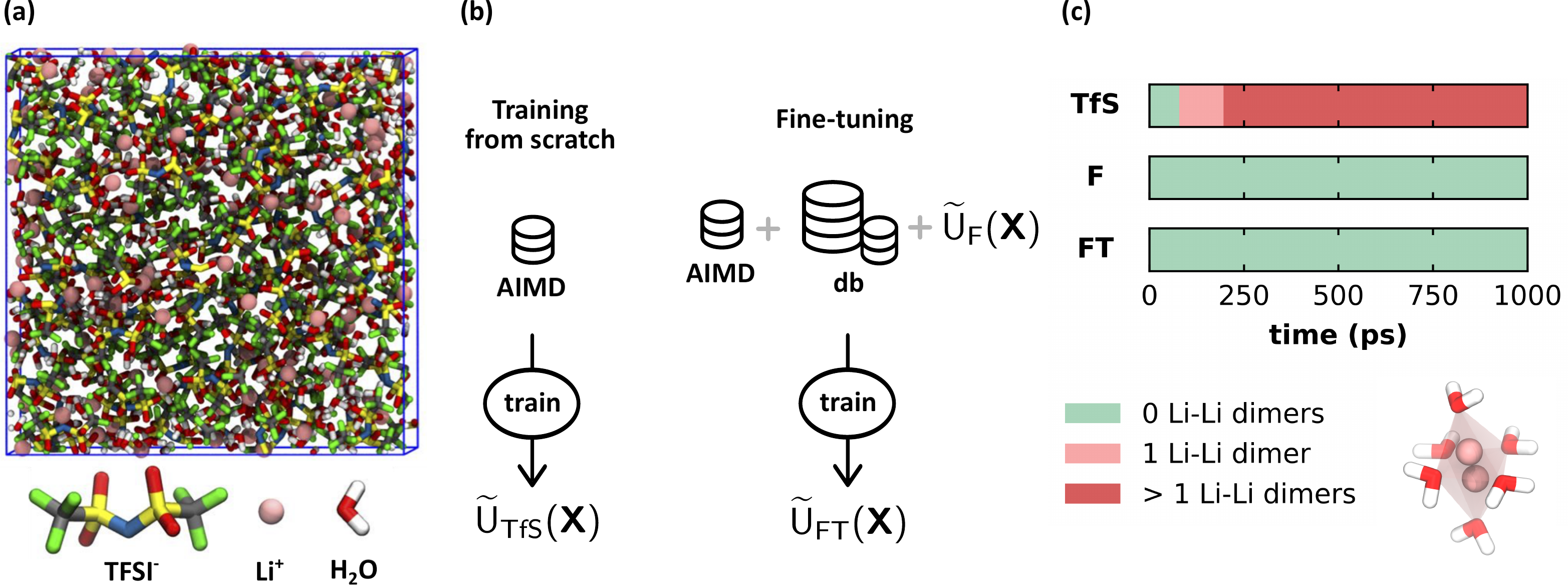}
    \caption{(a) Snapshot of the 20.9~m LiTFSI water-in-salt electrolyte. The bottom panel shows the molecular depictions used throughout: TFSI$^{-}$, Li$^{+}$, and H$_2$O; (b) Workflows of the two different training strategies, training from scratch (TfS) and fine-tuning (FT); (c) Occurrence of unphysical Li-Li dimers over a time span of 1 ns for three different models. A snapshot of such a dimer is displayed in the bottom panel. }
    \label{fig:intro}
\end{figure}

After training a series of models either from scratch (TfS) and through fine-tuning (FT) (as depicted in Figure \ref{fig:intro}b), we first evaluate their propensity compared to foundation (F) models to generate meaningful MD simulations of a 20.9 m LiTFSI WiSE (Figure \ref{fig:intro}a). We demonstrate that TfS models can lead to unphysical results due to datasets not covering rare configurations, such as short-range Li$^+$-Li$^+$ repulsion (see Figure \ref{fig:intro}c). Then, we move to investigating their capacity at reproducing the experimental density and diffusion coefficients, and discuss the effects of including corrections to dispersion interactions such as D3~\cite{Grimme2010D3}. Our results demonstrate that here, the choice of reference exchange-correlation (XC) functional is critical. After selecting four models based on these results, we then test these models on reproducing the experimental structure factor, compare to results obtained using short AIMD trajectories, and evaluate the influence of increased sampling time. This effect is also evaluated on the coordination of Li$^+$. This allows us to solve discrepancies between the AIMD and experimental structure factor: we show that converging features at low wavenumbers requires long trajectories. Finally, we evaluate the performance of different models in reproducing the experimental shear viscosity, for which significant sampling time is required. All together, our study offers a few key takeaways on using MLIPs to model WiSEs. First, foundation models should not be used out-of-the-box, as their performances are hard to anticipate. Second, we show the value of fine-tuning both for reducing the amount of data required for training, and as a simple solution to capture the correct physics in regions of configuration space hard to sample. Third, different XC functional lead to qualitatively different agreement with experimental data. Fourth, the inclusion of dispersion correction schemes should be validated. Finally, quantitative agreement with experimental data is achievable, but requires sampling times larger than what is currently reasonably reachable using AIMD; a corollary to this last point is that training on short AIMD trajectories can lead to MLIPs able to reproduce the correct experimental trends using subsequent long dynamics. Overall, our results show that carefully-designed MLIPs can accurately capture the physics of water-in-salt electrolytes. 

\section{Computational approach}

Our objective is to study the 20.9~m LiTFSI--H$_2$O mixture at 300~K; for that purpose, we used previously-reported AIMD trajectories as a starting point for dataset construction~\cite{Goloviznina2024Nanochains}. These trajectories were generated using a GGA XC functional (BLYP with D3 corrections). Here, we recompute energies and forces for the same set of configurations, at the meta-GGA level using the revised and regularised strongly constrained and appropriately normed (r$^2$SCAN) functional, which improves the description of hydrogen bonding and ion pairing in condensed aqueous solutions~\cite{Sun2016accurate,Chen2017ab,kaplanFoundationalPotentialEnergy2025}. We then benchmarked public MACE foundation models on this chemistry, and built task-specific models by both fine-tuning and training from scratch. Benchmarking of the various models was then achieved by computing a range of properties using MD simulations. 

\subsection{System composition and cells}

Two periodic and cubic boxes were used: a larger cell ("Li$_{128}$"), containing 128 LiTFSI pairs and 340 H$_2$O molecules (3068 atoms), and "Li$_{64}$", exactly twice smaller (1534 atoms). Both boxes are on the higher end of what is achievable using density functional theory. Unless specified, the density and diffusivity were evaluated using the Li$_{128}$ box, while viscosity was estimated from simulations performed using the smaller cell. Box vectors for initial configurations were set to reproduce the experimental density at $21 m$ of $1.7126$~g\,cm$^{-3}$\cite{gilbertDensityViscosityVapor2017}; volumes were subsequently allowed to relax in NPT runs (see below).

\subsection{Electronic structure reference data}

All electronic structure calculations were performed using CP2K/QUICKSTEP version 9.1~\cite{kuhne2020cp2k,VandeVondele2005QUICKSTEP}, which employs a hybrid Gaussian and plane wave (GPW) scheme. We used Goedecker-Teter-Hutter pseudopotentials~\cite{goedecker1996separable,hartwigsen1998relativistic,krack2005pseudopotentials} throughout. The SCF convergence criterion was set to $10^{-7}$~Ha. Single point calculations on the configurations extracted from the Li$_{64}$ and Li$_{128}$ trajectories of Goloviznina \textit{et al.} ~\cite{Goloviznina2024Nanochains} were performed using a molecularly optimized, TZV2P-GTH-SCAN basis set\cite{vandevondele2007gaussian}, and a plane wave expansion cutoff of 1200 Ry (relative density cutoff of 70 Ry), to obtain well-converged atomic forces. Configurations exhibiting SCF convergence issues were discarded, as well as those showing any atomic force component exceeding 10~eV\,\AA$^{-1}$. Unless stated otherwise, a total of 1245 configurations were used for FT and TfS procedures, comprising 926 configurations from Li$_{64}$ and 119 configurations from Li$_{128}$. The test set was constructed using an independent trajectory from the ones used for training, and consists of 200 Li$_{64}$ configurations.

To provide an electronic-structure reference independent of the trajectories used for dataset construction, we also performed a separate AIMD simulation of the Li$_{128}$ cell using the r$^2$SCAN functional with no dispersion correction. For this simulation, we used a DZVP-MOLOPT-SR-GTH basis set~\cite{vandevondele2007gaussian}, and plane wave cutoff of 900~Ry, (relative density cutoff of 70~Ry). MD simulations in the Born--Oppenheimer approximation were performed with a 0.5~fs time step, targetting the NVT ensemble at 300~K using a Nos\'{e}--Hoover thermostat with a 0.1~ps time constant. The starting configuration was identical to the pre-equilibrated structure used for all machine learning MD (MLMD) runs (see Section 1 of the SI). After 5~ps of equilibration, 15~ps of production trajectory were collected and used for structural analysis. 
At odds with what Goloviznina \textit{et al.} reported earlier~\cite{Goloviznina2024Nanochains}, no Li$^+$ nanochains form in this trajectory. We investigated this phenomenon using various XC functionals: results are discussed in Section S3 of the SI. 

\subsection{Machine-Learned Interatomic Potentials}

We employed MACE~\cite{batatiaMACEHigherOrder2023,kovacsEvaluationMACEForce2023}, an equivariant, many‑body message‑passing neural network that constructs atomic energies from high‑order (up to 4‑body) equivariant messages while enforcing $O(3)$ symmetry via spherical harmonics/Wigner D-matrices. The architecture achieves state‑of‑the‑art force accuracy with two message‑passing layers. As baselines, we screened publicly released MACE foundation checkpoints from the \model{mace-foundations} environment~\cite{MACEFoundationsGitHub,MACEdocsFoundationModels}. This covers the MPTrj-based \model{mace-mp-0} line (including the successive \model{0a/0b/0b2/0b3} releases), \model{mace-mpa-0} (trained on a larger dataset combining MPTrj and sAlex), \model{mace-omat-0} (OMAT dataset), and MatPES-based models at different reference levels (\model{mace-matpes-pbe-0} and \model{mace-matpes-r$^2$scan-0}). These families primarily differ by the training dataset and by the underlying electronic-structure level used for labels, namely PBE+U for \model{mp}/\model{mpa}/\model{omat}, PBE for \model{matpes-pbe}, and r$^2$SCAN for \model{matpes-r$^2$scan}. When available, multiple capacity variants (e.g., small/medium/large) were considered within a given family.
Within the MPTrj line, the \model{0a/0b/0b2/0b3} releases share the same overall training corpus and reference level but incorporate successive robustness and stability updates documented by the developers.

\textbf{Training  protocols.} All models were trained using an energy and force loss, with weights reported in the SI. We used a multi‑head replay strategy for fine‑tuning foundation models, and conventional train/validation splits (90/10) stratified by configuration source. Hyperparameters (cutoff radius, channels, $L_{\max}$, number of layers, learning rate schedule, batch size) and other details concerning the training are listed in Tables~S2-S3. We report the mean absolute error (MAE) and root mean square error (RMSE) for forces and energies; the primary selection metric is the force RMSE on the held‑out test set.

\textbf{Dispersion.} Semi-local DFT reference models and finite-cutoff MLIPs do not recover the long-range London dispersion tail by construction, and dispersion corrections (e.g., Grimme D3/D4) are therefore widely used in condensed-phase simulations to improve the description of both solid and liquid phases~\cite{Grimme2010D3,RuizPestana2018QuestWater}. In this work, dispersion was treated as a modeling choice: in addition to "bare" trajectories (the interactions being predicted using MACE only, trained on bare DFT data), we generated "D3-corrected" trajectories by adding a Grimme D3 term with Becke--Johnson damping (D3(BJ)) on top of the MACE model~\cite{Grimme2011D3BJ}. This choice was primarily pragmatic and aimed at consistency across the screened model set, since the same D3(BJ) overlay can be applied uniformly within our workflow. For r$^2$SCAN-based descriptions, dispersion-corrected variants exist (with D4 or rVV10 being often adopted), but dispersion is not guaranteed to yield systematic improvements. In fact, D3 and rVV10 corrections applied to SCAN were shown to lead to overbinding in hydrogen-bonded molecular systems~\cite{Brandenburg2016benchmark,Wiktor2017note,Hermann2018electronic}; accordingly, we do not assume \textit{a priori} that D3-corrected trajectories are universally closer to experiments than bare ones~\cite{Ehlert2021r2SCAND4,Belleflamme2023DCr2SCAN,Dasgupta2021DCSCAN}. Conversely, dispersion corrections have been reported~\cite{bertaniAtomicStructureNa4P2S72025,Pedone2025SilicateGlassDispersion} to improve densities and related properties for MLIPs trained on dispersion-free semilocal datasets in disordered ionic and glassy systems, motivating us to asses the  sensitivity of the results also for PBE-based foundation models.

\subsection{Machine learning molecular dynamics}

All MLMD simulations were performed using the LAMMPS simulation package~\cite{Thompson2022LAMMPS}. Dynamics were integrated using the velocity-Verlet scheme, with a time step of 0.5~fs. Temperature was controlled with a chain of Nos\'e--Hoover thermostats~\cite{Nose1984,Hoover1985}, using a damping constant of 50~fs. NPT simulations for density equilibration were performed for the Li$_{128}$ box at 300~K and 1~atm, using an isotropic Nos\'e--Hoover barostat with a damping constant of 0.5~ps. We used the MACE--LAMMPS interface~\cite{MACEdocsLAMMPS}, with TorchScript-exported model files; representative input templates and element mappings are provided in the SI.

The LAMMPS \verb|pair_style dispersion/d3| implementation~\cite{LAMMPSdispersionD3docs} of the D3 correction scheme was used. We used a global cutoff of 16 \AA, along with a coordination-number cutoff of 10 \AA.

\subsection{Structural analysis}

Structural observables were computed for a small set of models, and compared against \textit{ab initio} and experimental references where available. Partial radial distribution functions (RDFs, $g_{\alpha\beta}(r)$) were evaluated for key pairs, including Li--Li and Li--O$_\mathrm{tot}$, where O$_\mathrm{tot}$ denotes oxygen atoms from both water molecules and the TFSI$^{-}$ anion. Coordination numbers were obtained by integrating the corresponding RDFs up to the first minimum. Total X-ray structure factors, $S(q)$, were computed from the trajectories using a Debye-scattering-based implementation and standard X-ray form-factor weighting, as implemented in the TRAVIS program~\cite{BrehmKirchner2011TRAVIS}. To enable a direct comparison to short AIMD trajectories, structural properties for MLMD were evaluated both over the first 20 ps and over a later and longer time window (1--2 ns), while AIMD derived quantities were averaged over 15 ps, after the initial 5 ps of equilibration. Trajectory inspection and molecular visualization were performed with VMD~\cite{Humphrey1996VMD}. 

\subsection{Transport properties}

\textbf{Self-diffusion.} Self-diffusion coefficients were extracted from the long-time slope of the mean-squared displacement (MSD) computed from unwrapped trajectories. For TFSI and H$_2$O, the MSDs of the N and O atoms, respectively, were tracked. For each trajectory, an initial equilibration interval of 200~ps was discarded, and diffusion coefficients were fitted over the diffusive regime. Uncertainties were estimated by dividing each analysed trajectory segment ($\ge$1~ns) into five contiguous blocks of equal length and reporting the mean and block-to-block variability~\cite{FlyvbjergPetersen1989}. Finite-size effects were corrected using the Yeh--Hummer analytical equation, using the simulation cell length and the experimental viscosity (32~mPa$\cdot$s)~\cite{YehHummer2004,Suo2015WiSE,ding2017conductivity}. 

\textbf{Nernst--Einstein conductivity.} The Nernst--Einstein conductivity $\sigma_\mathrm{NE}$ was computed from the (finite-size corrected) self-diffusion coefficients of Li$^{+}$ and TFSI$^{-}$ using stoichiometric charges, as detailed in the SI.

\textbf{Viscosity.} Shear viscosity was computed from NVT trajectories using a Green--Kubo stress autocorrelation approach~\cite{Hess2002Viscosity,Maginn2019BestPracticesTransport}. For each selected model, 20~ns trajectories of the Li$_{64}$ box were analysed. The stress autocorrelation function was evaluated up to a maximum lag time of 1000 ps, and the resulting viscosity estimates were averaged over 5 stress-tensor components and block-averaged to estimate uncertainties (full details and convergence checks are given in the SI). 

\section{Results and discussion}

\begin{figure}[h!]
    \centering
    \includegraphics[width=\textwidth]{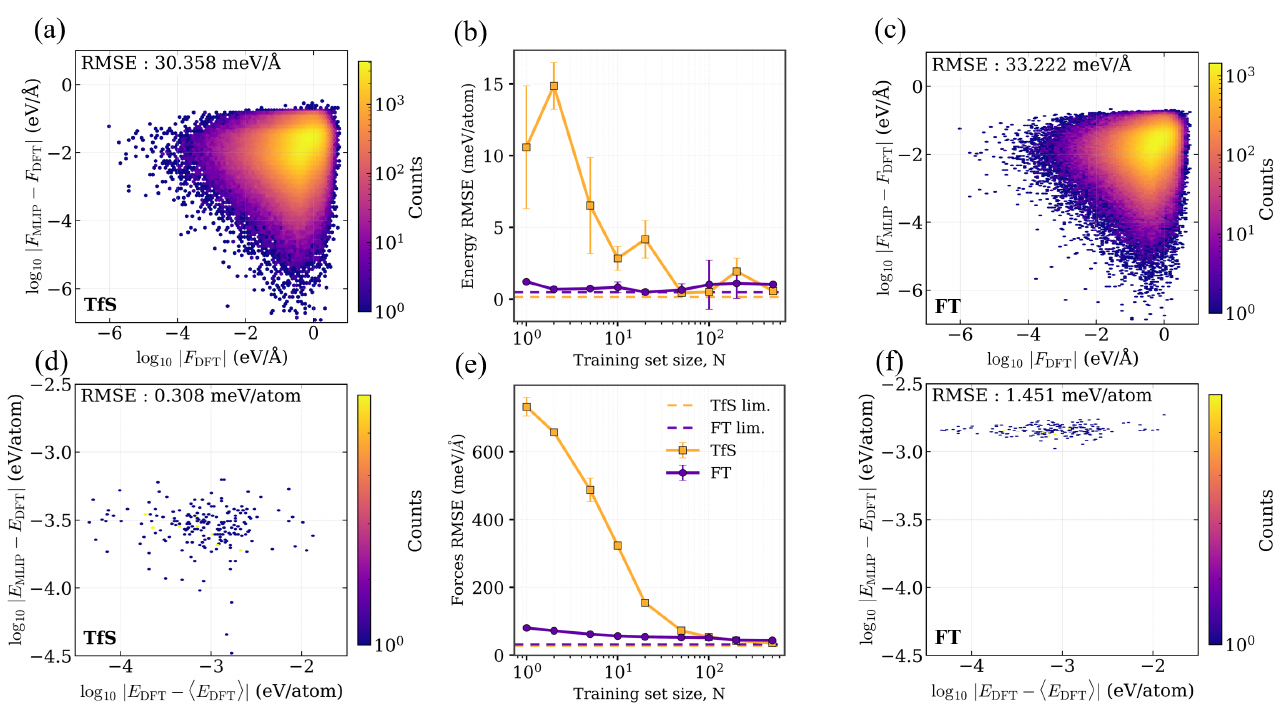}
    \caption{(a,d) Log--log density plots of absolute prediction errors for force components, and per-atom energies as a function of the corresponding r$^2$SCAN reference magnitude for the FT model, evaluated on the same  test set.  (b,e) Learning curves of RMSE versus training-set size $N$ for forces (b), and energies (e), comparing FT (purple) and TfS (orange). Points are averages over three independent replicas. Dashed lines indicate the RMSE obtained when using the largest training-set size in this sweep. (c,f) Same as (a,d), but for the TfS model.}
    \label{fig:errors_learning_curves}
\end{figure}

In Figure~\ref{fig:errors_learning_curves}, we compare FT and TfS results through error density maps (panels a,d, and c,f) and learning curves (panels b,e) for both atomic forces and energies. Both models share the same architecture (from the OMAT-medium foundation model), but FT starts from pretrained weights, whereas TfS starts from random initialization, highlighting the practical gain from pretraining in the case of this system. FT was initialized from the r$^2$SCAN-based checkpoint \model{mace-matpes-r2scan-omat-ft} to match the r$^2$SCAN flavour of the labels used in this work. Data-efficiency was assessed on Li$_{64}$-only datasets, using a fixed validation set (200 configurations) and training set sizes $N$ ranging from 1 to 500 configurations. For each $N$, three replicas were trained for FT and TfS using the same randomly drawn subsets (paired comparison), while the plotted points and error bars report the mean and replica-to-replica variability, respectively. In the small-data regime, the TfS force RMSE decreases steeply with increasing $N$, whereas FT starts at substantially lower force errors and improves more gradually (Figure~\ref{fig:errors_learning_curves}e). 

On the held-out test set, the FT model achieves RMSEs of 33.22~meV~\AA$^{-1}$ for forces and 1.451~meV~atom$^{-1}$ for energies (Figure~\ref{fig:errors_learning_curves}c,f), whereas the TfS model reaches 30.36~meV~\AA$^{-1}$ and 0.308~meV~atom$^{-1}$, respectively (Figure~\ref{fig:errors_learning_curves}a,d). The force error maps (Figure~\ref{fig:errors_learning_curves}a,c) show the absence of significant outliers in the predicted values. For energies, errors are plotted against $\left|E_\mathrm{DFT}-\langle E_\mathrm{DFT}\rangle\right|$, and TfS displays both a lower RMSE and a narrower vertical spread across the sampled range (Figure~\ref{fig:errors_learning_curves}b,f).  A key difference between the two protocols is the force loss weight: FT required $w_F = 200$ to effectively adapt the pretrained parameters, whereas in TfS, we set $w_F = 10$ (see SI, Tables~S2 and S3). This twenty-fold difference tilts the FT optimisation more strongly toward forces at the expense of energies, which probably accounts for the higher FT energy RMSE. As a consequence, the learning curves in Figure~\ref{fig:errors_learning_curves} compare training strategies, each with its respectively optimal hyperparameters, rather than architectures under a single fixed loss. In addition, the FT energy learning curve is non-monotonic at small-to-intermediate $N$ (Figure~\ref{fig:errors_learning_curves}b); we did not apply a two-stage schedule that reweights the loss toward energies once force convergence is approached, in order to keep the comparison as controlled as possible. The residual energy gap may also reflect the fact that FT retains parameter correlations from the pretraining domain, which can slow absolute energy calibration on a system-specific dataset. Beyond the Li$_{64}$-only learning curves, we also trained FT and TfS models on an expanded dataset combining 926 Li$_{64}$ configurations with 119 Li$_{128}$ configurations, and this increased diversity produced a further, modest reduction of test errors relative to the Li$_{64}$-only $N=500$ endpoint, without a qualitative change in the forces and energy error hierarchy. 

In short NVT MD test runs, unphysical Li–Li dimers at separations of 1.4–1.5 Å started to form using the TfS model trained exclusively on Li$_{64}$ data,  within the first few hundred picoseconds (Figure~\ref{fig:intro}c, "TfS" bar). Visual inspection of the dimers reveals that each Li–Li pair is embedded in a pseudo-octahedral cage of approximately six water molecules (Figure~\ref{fig:intro}c, bottom panel), suggesting that the artefact is stabilised by a cooperative effect: the potential energy surface develops a spurious attractive well for Li–Li at short range, while the surrounding Li--O(H$_2$O) coordination shell remains energetically reasonable, yielding an overall local minimum that traps the dimer once formed. Although analytical pair repulsion was active during training (\verb|pair_repulsion = True|), the universal ZBL potential for Li--Li ($Z = 3$) provides only modest repulsion at the relevant distances: approximately $1.3~\mathrm{eV}$ at $1.5~\text{\AA}$ and less than $0.25~\mathrm{eV}$ at $2.3~\text{\AA}$. This seems to be insufficient to compensate for the spurious attractive interaction in the MACE potential energy surface, possibly amplified by the favorable solvation shell. The origin of this effect can be deduced from the pair distance distributions of the training data (Figure S1): the Li$_{64}$ dataset contains no Li–Li distances below approximately 2.8~\AA{}, leaving the short-range interactions unconstrained during training from scratch. By contrast, the Li$_{128}$ dataset populates the $\sim 2.2-2.8$~\AA{} range, providing the seemingly necessary information that a TfS model requires to avoid the creation of such dimers. Including the Li$_{128}$ data in training eliminates the dimer artefact entirely. Most importantly, none of the FT models (including one trained on only 50 Li$_{64}$ configurations) exhibited Li–Li dimers: the pretrained foundation checkpoints tested seemingly provide information regarding some interaction ranges not covered in our AIMD datasets. In fact, when analyzing the distribution of Li-Li distances of one of these foundation datasets (\model{matpes}), short distances are covered (Figure S1). 

For the subsequent screening and production simulations we therefore retained two TfS variants, \model{tfs\_c128\_lmax1} (\model{omat}/\model{matpes}-like complexity) and \model{tfs\_c64\_lmax3} (reduced capacity with roughly halved channels and $l_{\max}=3$), together with \model{ft\_matpes-r2scan} and \model{ft\_matpes-r2scan\_s50}.
Having established the force and energy accuracy of the shortlisted models and verified that fine-tuning prevents the Li–Li dimer artifact across data regimes, we turn to the assessment of thermodynamic, structural, and transport properties extracted from MD simulations. 

\begin{figure}[h!]
\includegraphics[width=0.6\textwidth]{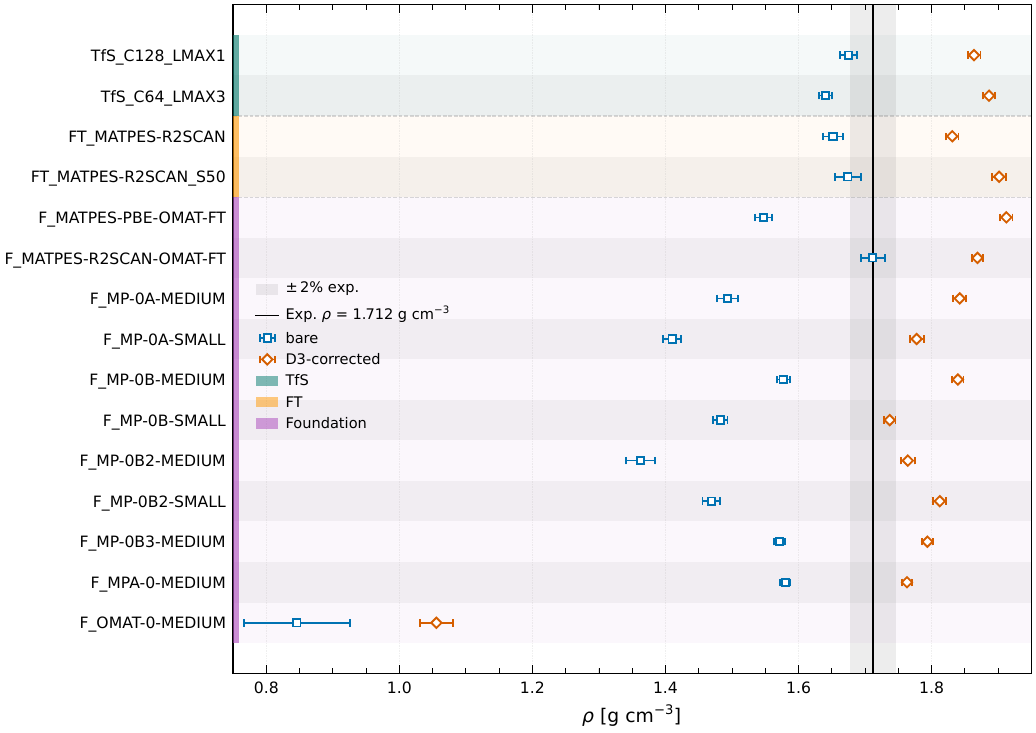}
\caption{Mean mass density $\rho$ (g\,cm$^{-3}$) of the 20.9~m LiTFSI WiSE. Results are shown for multiple MACE-based potentials, including FT, TfS, and foundation (F) models. Blue squares and red diamonds correspond to bare and D3-corrected models, respectively. Horizontal error bars indicate the estimated uncertainty from the finite sampling time. The vertical black line marks the experimental density.}
\label{fig:density_raw_d3}    
\end{figure}

Figure~\ref{fig:density_raw_d3} reports the equilibrium mass density obtained from 5~ns NPT trajectories of the Li$_{128}$ system, and provides a first assessment of model fidelity. Overall, bare models generally underestimate the experimental density, while D3-corrected ones systematically overshoot $\rho$, which is consistent with a more cohesive effective interaction. The \model{omat} foundation model is a clear outlier, yielding a substantially underestimated density even with a D3 correction, indicating a strong mismatch between that checkpoint and the present WiSE chemistry. The r$^2$SCAN-based \model{matpes} foundation checkpoint (\model{f\_matpes-r2scan-omat-ft}) reproduces the experimental density closely in the bare variant, while its D3-corrected counterpart overshoots, showing that an \textit{a posteriori} dispersion term is not automatically beneficial when the underlying short-range ML interaction is already consistent with an r$^2$SCAN reference. Fine-tuning this r$^2$SCAN foundation model on the WiSE dataset shifts the predicted density away from experiment (both for the full-data FT and for TfS), whereas the few-shot fine-tuned model (\model{ft\_matpes-r2scan\_s50}) remains closer to the foundation model estimate, consistent with partial retention of the foundation prior at low data. Overall, Figure~\ref{fig:density_raw_d3} highlights that comparable fitting errors do not guarantee an accurate equation of state, and motivates using density as an explicit down-selection criterion before structural and transport validation. 

\begin{figure}[h!]
\includegraphics[width=\textwidth]{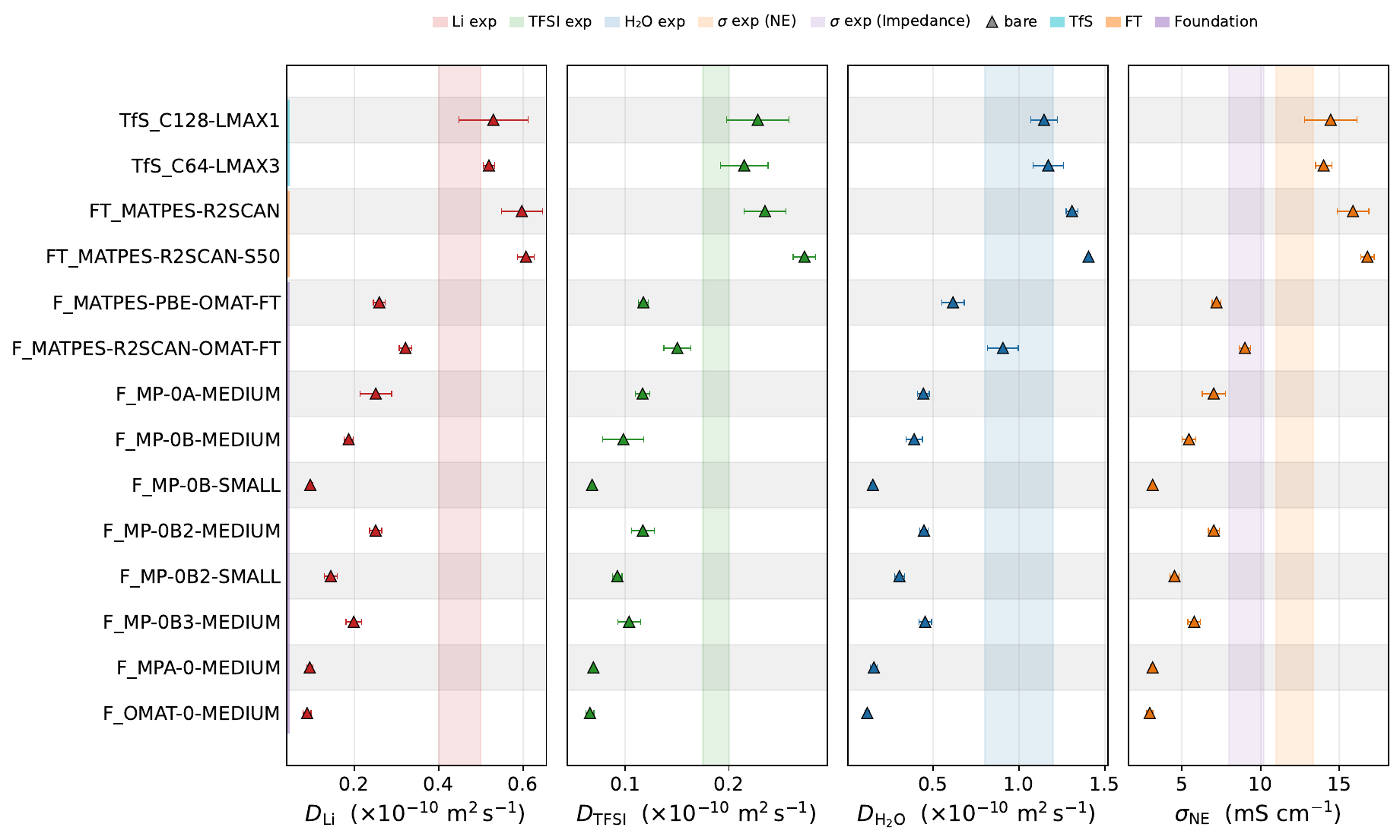}
   \caption{Transport properties computed using bare models. From left to right: diffusion coefficients ($D_{\mathrm{Li}}$, $D_{\mathrm{TFSI}}$, $D_{\mathrm{H_2O}}$), and Nernst-Einstein conductivity ($\sigma_{\mathrm{NE}}$). Shaded bands correspond to the experimental ranges.}
\label{fig:diffusion_conductivity_ne}
\end{figure}

Figure~\ref{fig:diffusion_conductivity_ne} summarizes transport properties estimates from NVT trajectories run at the experimental density, reporting self-diffusion coefficients and the corresponding Nernst--Einstein conductivity estimate $\sigma_\mathrm{NE}$. The Yeh-Hummer correction to the diffusion coefficients is $\Delta D = 5.62 \times 10^{-12}$~m$^{2}$\,s$^{-1}$, which is small compared with $D_{\mathrm{H_{2}O}}$ but amounts to 10--15\,\% of $D_{\mathrm{Li}^{+}}$ and can exceed 30\,\% for TFSI$^{-}$ in the experimental range. 

Experimental PFG-NMR self-diffusion coefficients for the 20--21~m LiTFSI WiSE have been reported by Borodin \textit{et al.}\cite{Borodin2017Nanoheterogeneity} (at 293~K) and by Li \textit{et al.}\cite{Li2019transport} (at 303~K); the shaded bands in Figure~\ref{fig:diffusion_conductivity_ne} bracket the ranges spanned by these two studies ($D_{\mathrm{Li}} \approx 0.4$--$0.5$, $D_{\mathrm{TFSI}} \approx 0.17$--$0.2$, and $D_{\mathrm{H_2O}} \approx 0.8$--$1.2$, all in units of $10^{-10}$~m$^{2}$\,s$^{-1}$). From these experimental diffusion ranges we derive a Nernst--Einstein conductivity estimate of $\sigma_{\mathrm{NE}}^{\mathrm{exp}} \approx 11$--$13$~mS\,cm$^{-1}$. The actual conductivity range, derived from impedance spectroscopy experiments, ($\sigma_{\mathrm{exp}} \approx 8$--$10$~mS\,cm$^{-1}$) collects measurements from Suo \textit{et al.}~\cite{Suo2015WiSE}, Ding \textit{et al.}\cite{ding2017conductivity}, and Borodin \textit{et al}~\cite{Borodin2017Nanoheterogeneity}. Consistent with our observations on density, most foundation models predict markedly slower ion dynamics (and therefore lower $\sigma_{\mathrm{NE}}$), whereas the WiSE-specific FT and TfS models cluster closer to the experimental diffusion ranges simultaneously for Li$^{+}$, TFSI$^{-}$, and H$_2$O, and mechanically yield $\sigma_{\mathrm{NE}}$ values nearer to the experimental Nernst-Einstein range. We note that $\sigma_{\mathrm{NE}}$ neglects ion--ion cross-correlations and therefore constitutes an upper bound to the true ionic conductivity; the gap between the NE-derived and impedance bands in Figure~\ref{fig:diffusion_conductivity_ne} (${\sim}30$\%) is consistent with the degree of ion correlation expected in highly concentrated WiSEs~\cite{ZhangMaginn2021WiSE2}. 

The transport properties calculated using D3-corrected models are reported in the SI (Figure S3). Comparing bare and D3-corrected variants does not reveal a single universal shift applicable to every model, although dispersion often seems to slows down diffusion by increasing cohesive interactions. Importantly, these transport comparisons are performed in NVT at a fixed, experimental volume, while different models (and their D3-corrected counterparts) exhibit different equilibrium densities (Figure~\ref{fig:density_raw_d3}). As a result, each model is effectively probed at a different internal pressure. Taken together, the combined density and transport validation tests motivate focusing subsequent experiments on structural properties (RDFs and X-ray $S(q)$) on the WiSE-adapted models (\model{ft\_matpes-r2scan}, \model{ft\_matpes-r2scan\_s50}, \model{tfs\_c64\_lmax3}, \model{tfs\_c128\_lmax1}), while retaining one foundation model (\model{F\_matpes-r2scan-omat-ft}) for comparison. 

\begin{figure}[h!]
\centering
\includegraphics[width=\textwidth]{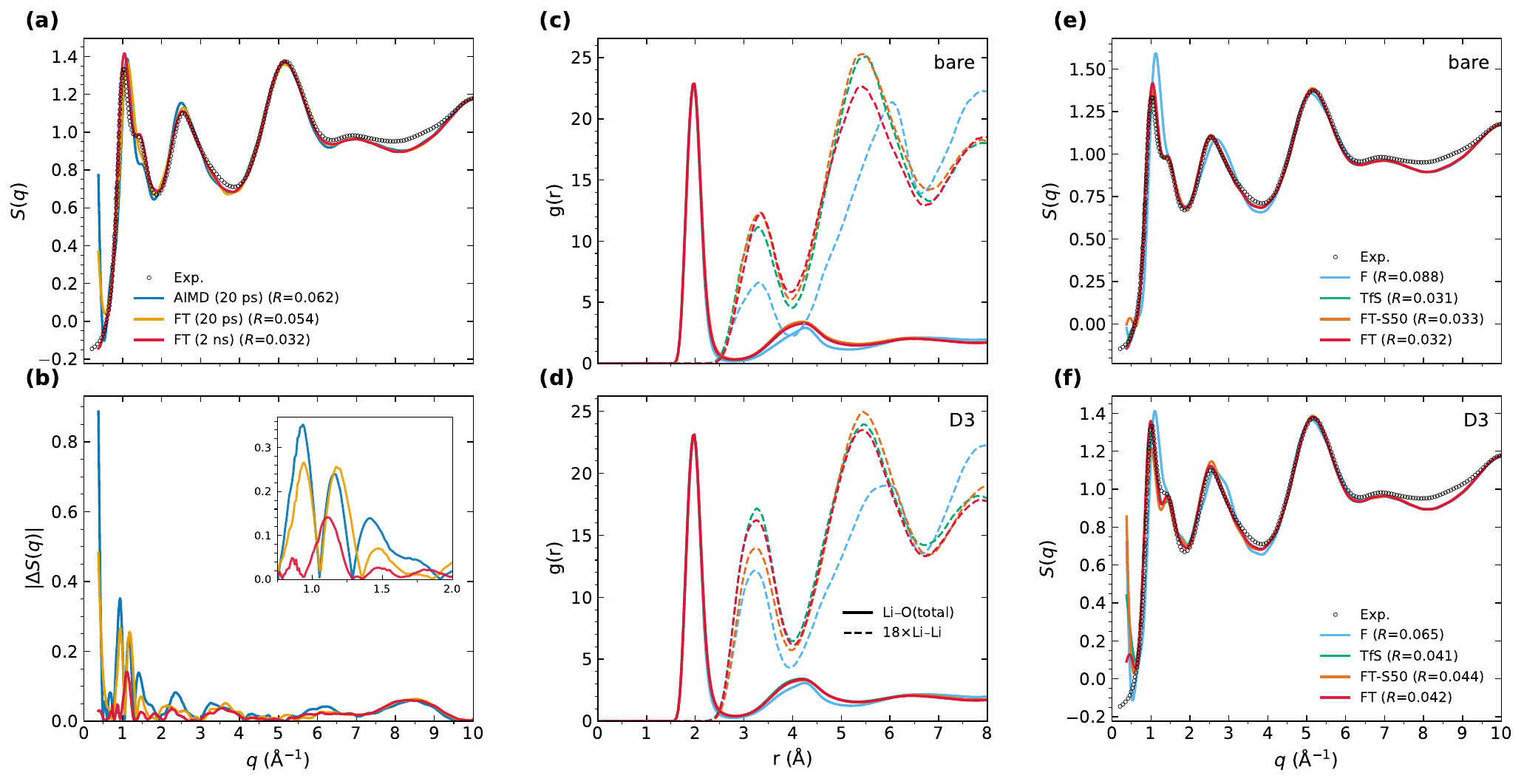}
\caption{Structural analysis. (a) Time- and model-dependence X-ray structure factor $S(q)$, comparing the experimental reference (black symbols, from Ref. \cite{zhangWaterinSaltLiTFSIAqueous2021a}) with estimates obtained from short (20 ps)  and long (2 ns) MD trajectories.Legend labels report the $R$-factor (defined in the text) for each computed curve. The abbreviation \textbf{FT} refers to the fine-tuned \mbox{FT\_MATPES-R2SCAN} model throughout.
(b) Absolute deviation $|\Delta S(q)| = |S_{\mathrm{sim}}(q)-S_{\mathrm{exp}}(q)|$ for the three simulated curves shown in (a); the inset magnifies the $q=1$--$2~\AA^{-1}$ region.
(c,d) Radial distribution functions (RDFs, $g(r)$) for Li--O(total) (solid lines, oxygen atoms from both H$_2$O and TFSI$^{-}$) and Li--Li (dashed lines, scaled by a factor of 18 for visibility), for (c) bare and (d) \textit{D3-corrected} trajectories. 
(e,f) Total X-ray $S(q)$ from long-time MLMD trajectories for four selected models compared to experiment: (e) bare, (f) \textit{D3-corrected} MLMD. Legend labels indicate the model abbreviation and the corresponding $R$-factor. 
In panels~(e) and~(f), \textbf{F} denotes the foundation model (\mbox{F\_MATPES-R2SCAN-OMAT-FT}), \textbf{TfS} the trained-from-scratch model (\mbox{TfS\_C128\_LMAX1}), \textbf{FT-S50} the few-shot fine-tuned model (\mbox{FT\_MATPES-R2SCAN\_S50}, 50~configurations), and \textbf{FT} the fully fine-tuned model (\mbox{FT\_MATPES-R2SCAN}).}
\label{fig:rdf_sq_comparison}
\end{figure}

We now turn our focus to structural validation through the X-ray structure factor $S(q)$ and selected radial distribution functions. In Figure~\ref{fig:rdf_sq_comparison}a, the experimental $S(q)$ (black points) is compared to three computational results estimated from trajectories initialized from the same configuration: a short, 20~ps AIMD simulation, and two simulations using the \model{ft\_matpes-r2scan} model (20~ps and 2~ns).
To quantify the agreement between computed and experimental structure factors we adopt the crystallographic $R$-factor, defined as

\begin{equation}
  R = \sqrt{
    \frac{\displaystyle\sum_{i} \bigl[S_{\mathrm{calc}}(q_i)
          - S_{\mathrm{exp}}(q_i)\bigr]^{2}}
         {\displaystyle\sum_{i} S_{\mathrm{exp}}(q_i)^{2}}
  }\,,
  \label{eq:rfactor} 
  \end{equation}

where the sums run over a uniform grid ($\Delta q = 0.05$~\AA$^{-1}$, $q \in [0.5, 10]$~\AA$^{-1}$) onto which both curves are interpolated; this ensures that every spectral region contributes with equal weight regardless of the original sampling density. 
The $R$-factor values are reported alongside each model label in the legends of panels~(a), (e), and~(f).
Features at wavenumbers greater than 2~\AA$^{-1}$ are well represented, and show only small variations. The two most discriminating features in this panel are the first maximum ($q \simeq 1$~\AA$^{-1}$), and the shoulder following it ($q \simeq 1.5$~\AA$^{-1}$). A pairwise decomposition of the structure factor, with comparison to experiments, attributes the first maximum to TFSI$^-$-TFSI$^-$ and TFSI$^-$-H$_2$O interactions~\cite{zhangWaterinSaltLiTFSIAqueous2021a}. The two simulations based on short dynamics give a first maximum at a too high wavenumber. For these simulations, the shoulder is either too low (AIMD), or too shallow (FT model). Longer sampling using the FT model resolves both issues: the peak is shift to the correct position (the intensity is still slightly too large), and the shoulder accurately matches the experimental signal. An analysis (Figure S9) shows that a box size of roughly 27~\AA (corresponding to 64 ion pairs) is needed to resolve this shoulder, consistent with its nanoscale origin. The low-wavenumber features ($q \to 0$) also show strong time-dependence; the results obtained with the FT model and 2 ns of sampling have an excellent agreement with the experimental data. These observations are consistent with the notion that slow structural relaxation and long-lived ion-association patterns can lead to biased time averages from short simulations in viscous or persistent solvated-ion environments, where equilibration and sampling strategy matter~\cite{Crabb2020AIMDsolvatedions}. In this context, the 20 ps AIMD curve in Figure~\ref{fig:rdf_sq_comparison}a is best read as a pseudo-reference starting from the same configuration rather than a fully converged, equilibrium estimate. In fact, there is a measurable memory effect from the equilibration procedure used (such as using a classical analytical potential for pre-equilibration dynamics) for MD short with respect to the inherent dynamics of the system~\cite{Crabb2020AIMDsolvatedions}, which reduces the value of such simulations for a comparison to experimental data. 

The four right-hand panels of Figure~\ref{fig:rdf_sq_comparison} compare, on long timescales (4 to 5 ns), the Li-Li and Li-O RDFs (panels c,d) and the structure factor (panels e,f) computed using the four selected models, either with or without D3-corrections. Across all models, $g_{\mathrm{Li-O(total)}}(r)$ exhibits a sharp first-shell peak near $r \approx 2$~\AA\ and a broad second-shell feature around 4~\AA, with only modest model-to-model variation in peak positions and heights. By contrast, the Li-Li correlations show visibly larger differences among models, particularly in the position and depth of the minima and in the relative intensity of the oscillations between about 3 and 7~\AA. In the corresponding bare $S(q)$ panel (Figure~\ref{fig:rdf_sq_comparison}c), the three trained models show remarkable agreement with the experimental reference, while \model{f\_matpes\_r2scan\_omat\_ft} displays slightly different peak amplitude pattern in the low-to-intermediate $q$ oscillations. 

Including D3 corrections (Figure~\ref{fig:rdf_sq_comparison}d,f) has a comparatively small effect on the Li-Li and Li-O RDFs, but a pronounced impact on the structure factor, particularly at low wavenumber. The first peak of the Li-O $g(r)$ remains nearly unchanged, and the overall oscillation pattern for Li-Li is preserved, but small shifts in the dashed curves indicate that dispersion can slightly affect ion-ion correlations. It is important to note however that since the Li-Li signal is very low, these effects, as well as variations across models for the bare case, could be due to sampling effects. Regarding the $S(q)$, a plausible explanation for the variations noted is that low-$q$ scattering probes large length scales and is therefore sensitive to long-wavelength density fluctuations and compressibility-like responses, which are precisely the degrees of freedom most affected by how long-range interactions are represented. Here the dispersion overlay is intrinsically long-range, but is applied with a finite cutoff (16~\AA\ in the MLMD simulations). Any truncation artefact can feed into the low-$q$ regime and become visually dominant, even if the local structure is mostly unchanged. Overall, these results suggest that for the present setup, bare models provide a more faithful representation of structural characteristics than D3-corrected ones. 

\begin{table}
\centering
\caption{Lithium first-shell coordination numbers (CN) split into water oxygen (O$_\mathrm{w}$) and anion oxygen (O$_\mathrm{TFSI}$), as well as O$_\mathrm{total}$, reported for bare and D3-corrected models. Long-window values correspond to the model set of Figure~\ref{fig:rdf_sq_comparison}(e,f); the last two rows quantify time-window sensitivity for the few-shot model \model{ft\_matpes-r2scan\_s50}}.
\label{tab:cn}

\begin{threeparttable}
\begin{tabular}{lcccccc}
\toprule
& \multicolumn{3}{c}{Bare} & \multicolumn{3}{c}{D3-corrected} \\
\cmidrule(lr){2-4} \cmidrule(lr){5-7}
Model & CN(O$_\mathrm{w}$) & CN(O$_\mathrm{TFSI}$) & CN(O$_\mathrm{tot}$) &
        CN(O$_\mathrm{w}$) & CN(O$_\mathrm{TFSI}$) & CN(O$_\mathrm{tot}$) \\
\midrule
AIMD (20 ps)$^{\dagger}$ & 2.23 & 2.02 & 4.31 & 2.36 & 2.01 & 4.37 \\
{\model{f\_matpes\_pbe\_omat\_ft}} & 1.93 & 2.12 & 4.05 & 2.15 & 1.97 & 4.12 \\
{\model{f\_matpes\_r2scan\_omat\_ft}} & 2.06 & 2.04 & 4.09 & 2.23 & 1.91 & 4.14 \\
{\model{tfs\_c128\_l1}} & 2.38 & 1.87 & 4.25 & 2.52 & 1.86 & 4.38 \\
{\model{ft\_matpes-r2scan}} & 2.31 & 1.94 & 4.26 & 2.47 & 1.94 & 4.41 \\
\midrule
{\model{ft\_matpes-r2scan\_S50}} short & 2.25 & 1.93 & 4.18 & 2.53 & 1.77 & 4.31 \\
{\model{ft\_matpes-r2scan\_S50}} long & 2.50 & 1.73 & 4.23 & 2.73 & 1.64 & 4.37 \\
\bottomrule
\end{tabular}
\begin{tablenotes}
\footnotesize
\item[$\dagger$] The AIMD reference is limited to 20 ps and is included only as a short-time electronic-structure benchmark
\end{tablenotes}
\end{threeparttable}
\end{table}

An analysis of the Li$^+$ coordination environment offers insights consistent with what was deduced previously. In Table~\ref{tab:cn}, we compare the coordination numbers (CN) of Li by oxygen (from water, from TFSI, and from both), for all models mentioned above, as well as for a PBE foundation baseline (\model{f\_matpes\_pbe\_omat\_ft}). Fine-tuning and training-from-scratch strategies bring CN(O$_\mathrm{tot}$) close to the 20 ps AIMD reference, but the coordination by the anion appears to be more sensitive. 
In the long-time comparison, the foundation models seem to underestimate CN(O$_\mathrm{tot}$), mainly through a lower CN(O$_\mathrm{w}$) contribution. A detailed comparison of the trained models with the reference AIMD is however pointless: as we can see in the lower panel of Table~\ref{tab:cn}, CN(O$_\mathrm{tot}$) can appear nearly converged on short timescales, while the first solvation shell speciation continues to drift on nanosecond-long sampling. From 20 ps to $\sim$5 ns, CN(O$_\mathrm{tot}$) changes only from 4.185 to 4.231 (bare) and from 4.307 to 4.373 (D3), yet the partition shifts markedly toward higher CN(O$_\mathrm{w}$) and lower CN(O$_\mathrm{TFSI}$). This suggests that exchange between water-coordinated and anion-coordinated motifs is slower to converge than the total first-shell occupancy in this concentrated, viscous electrolyte. Therefore, we cannot base our analysis of the performances of the models on agreement with the AIMD data. Ideally, we would compare the models with coordination numbers deduced from spectroscopy experiments, although extracting this information is challenging~\cite{Han2022electrolyte}. 

As a final test, we estimated the shear viscosity using three models: \model{f\_matpes\_r2scan\_omat\_ft}, \model{ft\_matpes-r2scan}, and \model{tfs\_c128\_l1}. The predicted viscosities can be directly compared to the experimental value~\cite{Suo2015WiSE,ding2017conductivity}, $\eta_{\mathrm{exp}} \approx 32$~mPa·s. For the bare models, we obtain $\eta = 48 \pm 16$~mPa·s (\model{f\_matpes\_r2scan\_omat\_ft}), $\eta = 34 \pm 10$~mPa·s (\model{ft\_matpes-r2scan}, the closest to the experimental value), and $\eta = 23 \pm 14$~mPa·s  (\model{tfs\_c128\_l1}). Within the statistical uncertainty, all three estimates are very close to the experimental value. 
A box-size check using the foundation model indicates that finite-size effects are within the statistical uncertainty at the present level of sampling: using the Li$_{128}$ box, we obtain $\eta = 48 \pm 17$~mPa·s, very close to the Li$_{64}$ estimate (see Table S4). For the same model, including the D3 correction increases the viscosity to $\eta = 105 \pm 20$~mPa·s, suggesting that the added long-range attraction overestimates cohesive interactions and slows down relaxation; this supports our previous findings. 

\section{Conclusions}

We benchmarked various MACE MLIPs for the 20.9~m LiTFSI WiSE by comparing pretrained foundation models, fine-tuned models, and models trained from scratch. Benchmarking was performed by comparing properties -- both structural and dynamics --, extracted from MD simulations, to experimentally determined values. 

Reference data was obtained using the r$^2$SCAN XC functional, by re-evaluating energies and forces of configurations extracted from pre-existing AIMD trajectories. A central result is that fine-tuning provides robustness beyond what is captured by energy and force error metrics alone. While FT and TfS errors approach a similar plateau in the high-data regime, TfS models trained only on Li$_{64}$ configurations seem to be able to only poorly extrapolate in regions of configuration space not sampled. This for instance leads to the formation of unphysical Li$^+$--Li$^+$ dimers at distances lower than 1.6 \AA. In contrast, FT models inherit from the foundation datasets they were pre-trained on, which, even for models fine-tuned with a handful of configurations, leads to robustness with respect to such situations. This difference in robustness becomes decisive when moving from error metrics on held-out sets to long MD trajectories. In this sense, fine-tuning allows to overcome an important limitation in the sampling capacity of AIMD. 

Over all the benchmarks considered, the three trained models we test all the way (a TfS model, and two FT models, either in the low- or high-data regimes) perform almost similarly. In particular, they are all able to reproduce the experimental structure factor admirably well. This shows that one can obtain a very accurate model by fine-tuning from a small set of reference structures. 

Here, including a D3 analytical correction for dispersion interactions leads to a systematic worsening of the agreement with the experiments. We want to stress that this is certainly due to relying on r$^2$SCAN training data, a meta-GGA; we would probably not have obtained such results with GGA reference data. These results however show that one should carefully check for the choice of reference functional, and for the effect of any correction added \textit{a posteriori}. 

These MLIP models allow to overcome limited time-sampling inherent to AIMD. For highly viscous liquids such as WiSEs, this is an important restriction. For instance, it is hard to tell \textit{a priori} if the disagreements between the experimental structure factor and its estimation from short AIMD trajectories, both reported here or in the literature~\cite{Singh2025structure}, are due to insufficient sampling, or to inaccurate physical models. Using MLIPs, we are able to answer this question, showing that the correct experimental signal can be recovered from nanosecond-long MLMD trajectories. 

This becomes more critical when an experimental reference is lacking: for instance, regarding the Li$^+$ speciation. Our results show that estimates obtained from short and long timescales differ; while the overall coordination by oxygen is consistent with short-time AIMD estimates, the actual partitioning between water and TFSI changes. Properly equilibrating viscous electrolytes or tightly-bound solvation shells is a known problem~\cite{Crabb2020AIMDsolvatedions}, and MLIPs allow to reach simulation times for which memory from the structure generation method or from the classical potential used for pre-equilibration is lost. This strategy can be applied in future works to study other relevant families of ion concentrated
liquids, such as localized high-concentration
electrolytes or high entropy electrolytes that have recently been introduced.

\begin{acknowledgements}
The authors thank Rocio Semino for fruitful discussions  and Kateryna Goloviznina for providing AIMD simulation results. This work was granted access by GENCI to the HPC resources of CINES, IDRIS, and TGCC under the allocations A0160901387, A0170814184, and A0180901387. This work was supported by the French National Research Agency under the France 2030 program (Grant No. ANR-22-PEBA-0002), and by the by the Ile-de-France Region in the framework of Domaine d'Intérêt Majeur DIM MaTerRE. 
\end{acknowledgements}

\section*{Data and Software Availability}
Trained models, input scripts, and raw data will be uploaded upon publication of this manuscript. 

\section*{Notes}
Conflicts of interest: none.

\bibliography{refs}

\end{document}